\documentclass{aa}
\usepackage[dvips]{graphicx}

\newcommand{\ovi}{{\rm O}{\sc vi}\ }

\newcommand{\temp}{{\rm T}}
\begin{document}

   \title{The History and Future of the Local and Loop I Bubbles}
\subtitle{}

   \author{Dieter Breitschwerdt\inst{1}
          \and
          Miguel A. de Avillez\inst{1,2}
          }
   \offprints{D. Breitschwerdt}

   \institute{Institut f\"ur Astronomie, Universit\"at Wien,
     T\"urkenschanzstra{\ss}e 17, A-1180 Wien, Austria \\
        email: breitschwerdt@astro.univie.ac.at
     \and
     Department of Mathematics, University of \'Evora,
              R. Rom\~ao Ramalho 59, 7000 \'Evora, Portugal \\
              email: mavillez@galaxy.lca.uevora.pt
            }

   \date{Received 9 February, 2006; accepted 30 March, 2006}

   \titlerunning{Evolution of Local and Loop I Bubbles}
   \authorrunning{D. Breitschwerdt and M.A. de Avillez}

\abstract {The Local and Loop I superbubbles are the closest and
best investigated supernova (SN) generated bubbles and serve as test
laboratories for observations and theories of the interstellar
medium.}
{Since the morphology and dynamical evolution of bubbles depend on the
ambient density and pressure distributions, a realistic modelling of
the galactic environment is crucial for a detailed comparison with
observations.}
{We have performed 3D high resolution (down to 1.25 pc on a kpc-scale
grid) hydrodynamic simulations of the Local Bubble (LB) and the
neighbouring Loop I (L1) superbubble in a
\emph{realistically evolving inhomogeneous background ISM},
disturbed already by SN explosions at the Galactic rate for 200 Myr
before the LB and L1 are generated. The LB is the result of 19 SNe
occurring in a moving group, which passed through the present day
local H{\sc i} cavity.}
{We can reproduce (i) the \ovi column density in absorption within
the LB in agreement with \textsc{Copernicus} and recent
\textsc{FUSE} observations, giving ${\rm N}_{OVI }<2 \times 10^{13}
\, {\rm cm}^{-2}$ and ${\rm N}_{OVI }<7 \times 10^{12} \, {\rm
cm}^{-2}$, respectively, (ii) the observed sizes of the Local and
Loop~I superbubbles, (iii) the interaction shell between LB and L1,
discovered with ROSAT, (iv) constrain the age of the LB to be
$14.5\pm^{0.7}_{0.4}$ Myr, (v) predict the merging of the two
bubbles in about 3 Myr, when the interaction shell starts to
fragment, (vi) the generation of blobs like the Local Cloud as a
consequence of a dynamical instability.}
{We find that evolving superbubbles strongly deviate from idealised
self-similar solutions due to ambient pressure and density
gradients, as well as due to turbulent mixing and mass loading.
Hence, at later times the hot interior can break through the
surrounding shell, which may also help to explain the puzzling
energy ``deficit'' observed in LMC bubbles.}

\keywords{Hydrodynamics -- Shock Waves -- ISM: general -- ISM:
bubbles -- ISM: structure -- ISM: kinematics and dynamics}
\maketitle

\section{Introduction}

Our solar system is embedded in an H{\sc i} cavity extending about
200 pc in the Galactic plane and roughly 600 pc perpendicular to it,
according to Na{\sc i} absorption line studies towards background
stars (Lallement et al. 2003). For a long time it has been known
that at least part of this so-called Local Bubble (LB) shows diffuse
emission in soft X-rays (cf.\ McCammon \& Sanders 1990). Ever since
the LB has been the subject of intensive studies in radio, UV, EUV
and soft X-rays, as well as of analytical and numerical modelling
(e.g.\ Cox \& Anderson 1982, Breitschwerdt \& Schmutzler 1994, Smith
\& Cox 2001). However, due to the wealth of data and the inherent
complex structure of the LB, all models have severe shortcomings,
concerning either the dynamical evolution or the model spectra, in
some cases both. In particular models based on thermal conduction
usually fail to reproduce the observed low O{\sc vi} absorption
column density of $\sim 1.6\times 10^{13}$ cm$^{-2}$ in the Local
Bubble as inferred from a reanalysis of \textsc{Copernicus} data
(Shelton \& Cox 1994). One severe restriction in all the models,
which consider the LB to be the result of SN explosions, viz. the
expansion into a homogeneous ambient medium, has been removed in the
simulations presented here. As a corollary, the LB evolution has to
be performed jointly with that of the Loop~I (L1) superbubble. ROSAT
PSPC observations, revealing a coherent X-ray shadow towards L1, led
to the suggestion (Egger \& Aschenbach 1995) that the LB is in close
contact with the expanding neighbouring L1 superbubble. L1 thus
originated from multi-supernova explosions in the Sco Cen cluster,
with the North Polar Spur being part of its shock illuminated outer
shell.

It has also been suggested that part of the soft X-ray emission
could be of very local origin, generated by charge exchange
reactions between solar wind ions and heliospheric (e.g.\ Cravens
2000) or even atmospheric plasma. However, even in the most
favourable case of \emph{all} emission being \emph{very local} in a
certain direction, there is always missing flux along other lines of
sight, in particular perpendicular to the Galactic plane.

In this letter we describe for the first time a \emph{realistic}
evolutionary scenario for the \textit{origin and evolution} of the
Local Bubble jointly with the L1 superbubble. The crucial
differences as compared to previous models are: (i) a realistic
background medium, that has been pre-structured by previous
generations of SN explosions, which is therefore very inhomogeneous,
both in density and pressure distribution, (ii) taking into account
the time sequence and locations of SN explosions, which generate the
two bubbles, calculated according to the mass distribution of
extinct members of a moving group and the Sco Cen cluster,
respectively, both derived from a Galactic initial mass function
(IMF) normalised to still existing low mass stars, and (iii) to
calculate the density and temperature structure of the LB and L1 in
3D with high resolution in order to sample \ovi along many different
lines of sight and compare them to observations.

This letter is organised as follows: Section~2 describes details of
the model setup, in Section~3 our results with respect to bubble
evolution and comparison to observations are discussed, and we close
with a general discussion and conclusions in Section~4.
\begin{figure}[h]
\centering
\vspace*{4.2cm}
Jpeg image 4989f1a.jpeg\\
\vspace*{8cm}
Jpeg image 4989f1b.jpeg
\vspace*{4.2cm}

\caption{ \emph{Top:} Colour coded temperature map in the
range $10 \leq T \leq 10^7$ K for a slice through the data cube
(representing the Galactic midplane) of a 3D LB high resolution
simulation representing the present time (i.e.\ 14.5 Myr after the
first explosion) with the LB centred at (175, 400) pc and L1 shifted
200 pc to the right. \emph{Bottom:} Same, showing the ``future'' of
the LB and L1 at $t=29.7$ Myr. Note that part of the LB has merged
with the ISM, part has been engulfed by L1.}
\label{fig1}
\end{figure}

\section{Model and Simulations}
\label{mod}
We use a 3D parallelised adaptive mesh refinement (AMR) hydrocode to
track small scale structures down to 1.25 pc, where necessary, and we
follow the LB and L1 evolution in an ISM, driven by SNe types Ia, Ib+c
and II occurring at the Galactic rate (taken from Cappellaro et
al. 1999), on a Cartesian grid of $0\leq (x,y)\leq 1$ kpc size in the
Galactic plane, and $-10\leq z \leq 10$ kpc perpendicular to it. This
is an extension of the SN-driven ISM model of Avillez (2000), fully
tracking the time-dependent evolution of the large scale disk and the
Galactic fountain. The gravitational field of the stellar disk is
adopted from Kuijken \& Gilmore (1989).  Further processes include
radiative cooling, assuming an optically thin gas in collisional
ionisation equilibrium (CIE, i.e.\ ionisation rate by collisions
between electrons and ions is balanced by radiative recombination
rate) using the cooling functions of Sutherland \& Dopita (1993) for
$\temp\geq 10^{4}$ K and Dalgarno \& McCray (1972) for $\temp< 10^{4}$
K, with a temperature cut-off at 10 K, and uniform heating due to
starlight varying with height $z$ (Wolfire et al. 1995).  Galactic OB
associations form in regions with density $n\geq 10$ cm$^{-3}$ and
temperature T$\leq 100$~K, with the number, masses and main sequence
life times of the stars in the association being determined from the
IMF. All OB stars are allowed to drift away from their parent
associations with random velocities assigned according to
observations, before exploding either in clusters or in the field with
a canonical energy of $10^{51}$ erg.

The LB itself is carved by 19 successive explosions of the subgroup B1
of Pleiades, of which some low mass stars are now Sco Cen cluster
members according to the model of Bergh\"ofer \& Breitschwerdt (2002,
henceforth BB02). The adjacent L1 bubble is the result of explosions
of Sco Cen cluster members (see Egger 1998). Applying an IMF for OB
associations (Massey et al. 1995) and following BB02, we find that
about 39 SNe have occurred in L1 until now, and 38 SN candidates are
expected to explode within the next 13 Myr. In contrast, the LB is
extinct now, the last explosion having occurred about 0.5 Myr
ago. Main sequence life times are calculated from Stothers (1972).

\begin{figure*}[htb]
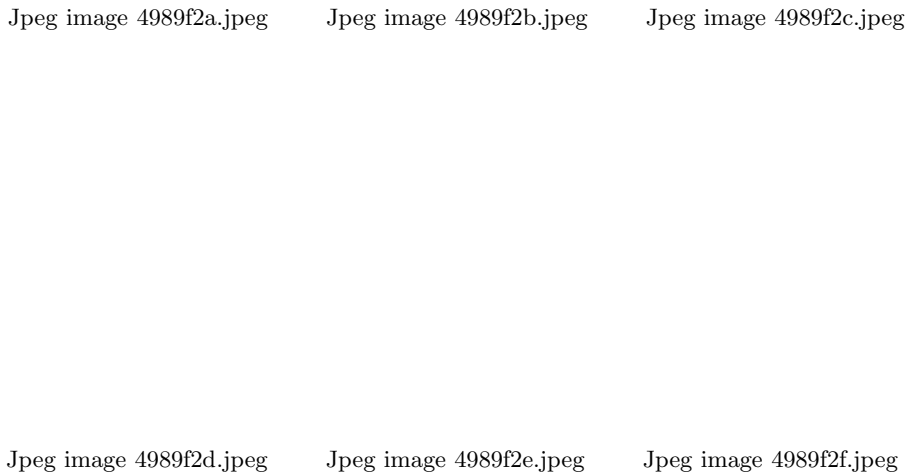

\centering
\vspace*{2.5cm}
Jpeg image 4989f2a.jpeg \qquad Jpeg image 4989f2b.jpeg \qquad Jpeg image 4989f2c.jpeg\\
\vspace*{5.5cm}
Jpeg image 4989f2d.jpeg \qquad Jpeg image 4989f2e.jpeg \qquad Jpeg image 4989f2f.jpeg
\vspace*{2.5cm}

\caption{Temperature distribution in the Galactic plane in
the range $10^5 \leq T \leq 10^6$ K of the LB and L1 region,
\emph{top:} at $t=14.5$ (left), $t=14.6$ (middle) and $t=14.7$ Myr
(right), and \emph{bottom:} at $t=14.8$ (left), $t=14.9$ (middle)
and $t=15.0$ Myr (right), after the first LB explosion. Note that in
L1, SN explosions are still occurring (see bottom panel), while the
LB is extinct.} \label{fig14.6to15.0}
\end{figure*}
As initial conditions for the present run we used data cubes of an
evolved SN driven ISM, in which global dynamical equilibrium has
been established (cf. Avillez \& Breitschwerdt 2004 (henceforth
AB04), 2005a). As the ISM structure reveals a typical pattern size
of a few hundred pc (Avillez \& Breitschwerdt 2005b, henceforth
AB05), we chose arbitrarily a site with a sufficient amount of cold
HI mass to form the Sco Cen cluster, here located at $(375,400)$ pc
(cf.\ Fig.~\ref{fig1}).  Then we followed the trajectory of the
Pleiades moving group B1, whose SNe in the LB went off along a path
crossing the location at $(175,400)$ pc, chosen to match the present
observed distances from the Sun and the Sco Cen. Due to small
peculiar motions of the local gas, the LB will be tied to the local
standard of rest (LSR; for details of the moving group kinematics
see BB02). Periodic boundary conditions are applied along the four
vertical boundary faces, while outflow boundary conditions are
imposed at the top ($z=10$ kpc) and bottom ($z=-10$ kpc) boundaries.
The simulation time of the LB and L1 evolution run was an additional 30 Myr.

\section{Results}
\label{results}

\subsection{Evolution of the Local and Loop I Superbubbles}
\label{lbev} The present runs were started after global dynamical
equilibrium, specifically the Galactic Fountain, had been
established (i.e. after $200$ Myr of evolution for a Galactic SN
rate; for details, see AB04). The locally enhanced SN rates produce
coherent LB and L1 structures within a highly disturbed and still
evolving background medium (due to ongoing star formation).
Successive explosions heat and pressurise the cavities, which at
first look smooth, but develop internal structure after some time;
for the LB this occurs at $t>8$ Myr after the first explosion. After
$\sim 14.5$ Myr (present time) 19 SNe have exploded inside the LB
cavity, which becomes progressively elongated and reaches a size of
180 by 220 pc in the Galactic midplane (Fig.~\ref{fig1} top), due to
the increasing influence of ambient density and pressure gradients.
Although magnetic fields and cosmic rays have an influence on bubble
evolution by tension and pressure forces, it seems unlikely that our
results will change significantly, because in the LB the last SN
went off only $0.5$ Myr ago, and Loop I is still active. If the
bubble pressure decreases markedly there may however be some
deviations.
Both bubbles are bounded by shells whose interaction already
generates Rayleigh-Taylor instabilities due to a larger pressure in
L1 with respect to the LB, in agreement with a linear stability
analysis of Breitschwerdt et al. (2000). After becoming nonlinear
this will lead to the formation of cloudlets, which will travel
towards the solar system from the Sco Cen region, in agreement with
observations of local clouds with measured velocities of about $26$
km/s. On the basis of our simulations we further predict that the
interaction shell will break up in $3$ Myr from now, allowing mass
transfer of hot gas from L1 to the LB, and in $\sim 15$ Myr from now
the bubbles will have merged (see Fig.~\ref{fig1} bottom).

In CIE, O{\sc vi} is most abundant at $T \sim 3 \times 10^5$ K,
whereas soft X-Ray emission of a thermal plasma requires a
temperature of $\sim 10^6$ K. Fig.~\ref{fig14.6to15.0} shows the
temperature distribution in the critical range, $10^5 \leq T \leq
10^6$ K, during the critical time, $14.5 \leq t \leq 15.0$ Myr, when
the LB plasma starts cooling down. It is obvious that between $14.5$
and $14.7$ Myr, cooler (O{\sc vi} absorbing) and hotter (X-ray
emitting) gas \textit{co-exist}. This is a direct result of the
turbulent and inhomogeneous density and temperature structure in
realistically evolved superbubbles. Turbulent mixing generates gas in
the whole thermally unstable regime between $10^5 - 10^6$ K.  While
the amount of O{\sc vi} in the LB has been measured fairly accurately
(see Sect.~\ref{ovi}), the source of the soft X-ray emission is more
elusive. First of all, it is more difficult to localise, as there are
most likely no dense clouds inside the LB, which can be used for
shadowing experiments, thus allowing us to separate back- and
foreground emission. Secondly, a yet uncertain amount of soft X-rays
will be produced locally by solar wind charge exchange reactions with
heliospheric plasma (Cravens 2000). And thirdly, the LB plasma need
not be in CIE. It has been shown that if a plasma undergoes fast
adiabatic expansion its kinetic temperature will drop, but
recombination can not follow fast enough, thus mimicking an
``overionised'' plasma (similar to the solar wind), where
recombination is delayed. In essence, the gas could be at a few times
$10^5$ K, but still emit copiously in soft X-rays (Breitschwerdt \&
Schmutzler 1994). We are currently working on these non-equilibrium
ionisation (NEI) models, and will present results in a forthcoming
paper.

\subsection{O{\sc vi} distribution in the solar neighbourhood}
\label{ovi} We have determined the amount of O{\sc vi} along the
lines of sight (LOS) for an ISM plasma in CIE with solar abundances
(Anders \& Grevesse, 1989; for details see AB05). To measure the
O{\sc vi} column density distribution inside LB and L1, we took 91
LOS extending from the Sun and crossing L1 (the hot pressured region
200 pc to the right of the LB) from an angle of $-45^\circ$ to
$+45^\circ$ as marked in the top panel of Fig.~\ref{fig1}.
\begin{figure}[htbp]
\centering
\includegraphics[width=0.6\hsize,angle=-90]{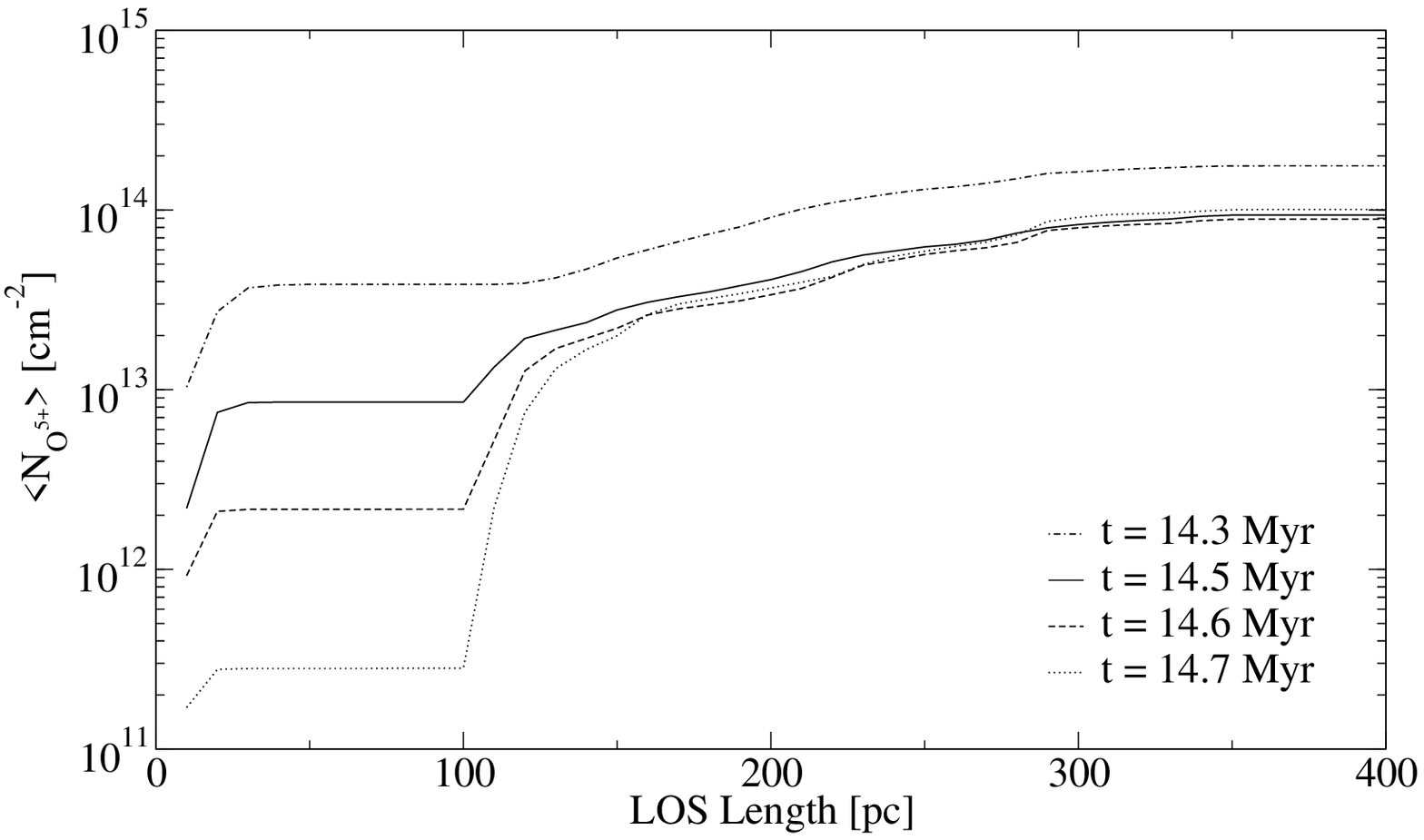}
\includegraphics[width=0.6\hsize,angle=-90]{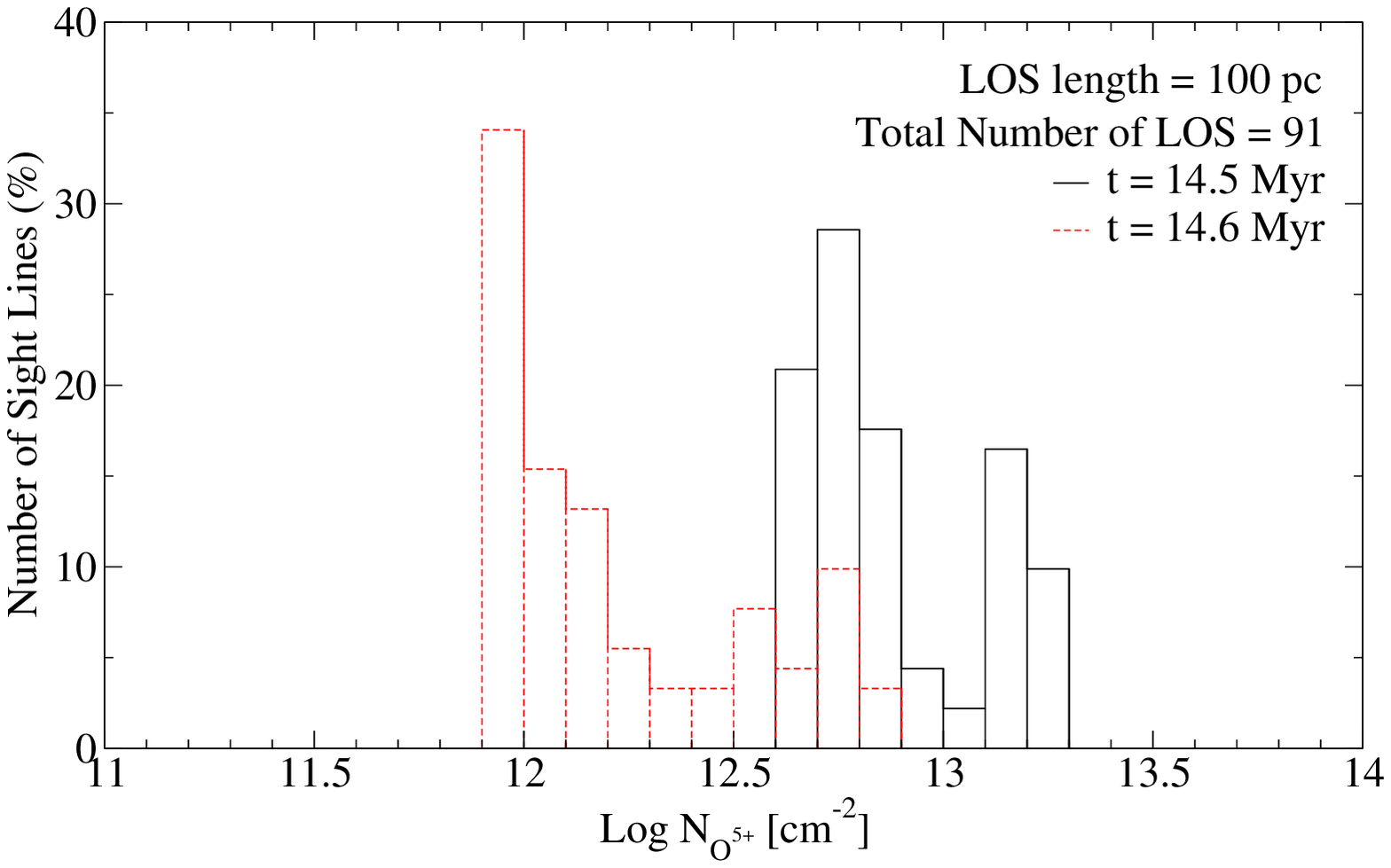}
\caption{ \textit{Top:} Averaged O{\sc vi} column densities (over 91 LOS between $\pm 45^\circ$ as marked in
Fig.~\ref{fig1} top) as a function of LOS path length at $14.3 \leq
t\leq 14.7$ Myr of Local and L1 bubbles evolution. \textit{Bottom:}
Histogram of the percentage of LOS as a function of $\mbox{N(O{\sc
vi})}$ within the LB at $t=14.5$ and $14.6$ Myr.} \label{figmaxave}
\end{figure}
Fig.~\ref{figmaxave} (top) shows the history (for $14.3\leq t\leq
14.7$ Myr) of the variation of the LOS averaged O{\sc vi} column
densities, $\langle\mbox{N(O{\sc vi})} \rangle$, with distance along
these LOS. They decrease steeply with time after the last SN in the
LB has exploded, from $4\times10^{13}$ to $3\times 10^{11}$
cm$^{-2}$, because no further heating is taking place. For LOS
sampling gas from outside the LB (i.e., $l_{LOS}>100$ pc) we have
$\langle \mbox{N(O{\sc vi})} \rangle > 2\times 10^{13}$ cm$^{-2}$.
The histograms of column densities obtained in the 91 LOS for
$t=14.5$ and $14.6$ Myr (Fig.~\ref{figmaxave}, bottom) show that for
$t=14.6$ Myr all the LOS have column densities smaller than
$10^{12.9}$ cm$^{-2}$, while for $t=14.5$ Myr 67\% of the lines have
column densities smaller than $10^{13}$ cm$^{-2}$ and about 50\% of
the lines have $\mbox{N(O{\sc vi})}\leq 7.9\times 10^{12}$
cm$^{-2}$. The maximum column density is $1.78\times10^{13}$
cm$^{-2}$ in excellent agreement with the observations discussed in
Savage \& Lehner (2006), which is 1.1 times the average value of
$1.6\times 10^{13}$ cm$^{-2}$ inferred by Shelton \& Cox (1994) from
a reanalysis of \textsc{Copernicus} absorption line data (cf.\
Jenkins 1978).

\begin{figure}[h]
  \centering
\includegraphics[width=0.9\hsize,angle=0]{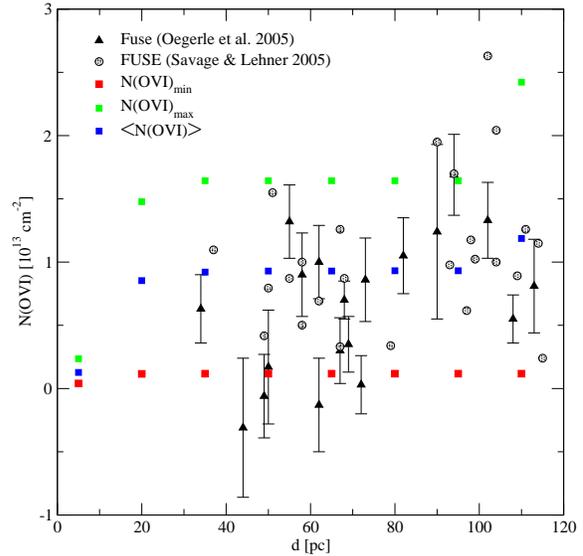}\\
\caption{
Comparison between FUSE (\emph{triangles:} Oegerle et al. 2005;
\emph{circles:} Savage \& Lehner 2006) O{\sc vi} column densities
with the calculated minimum, maximum and average column densities
along the same 91 lines of sight (at $t= 14.5$ Myr) pointing towards
L1 as shown in Fig.~1 (top). Negative values on the ordinate derive
from essentially zero equivalent absorption line widths, i.e.
non-detections, due to the noise in the spectrum (cf. Oegerle et al.
2005).
\label{FuseOegerle}}
\end{figure}

In Fig.~\ref{FuseOegerle} we compare FUSE data by Oegerle et al.
(2005; triangles) and Savage \& Lehner (2006; circles) with
simulated minimum (red squares), maximum (green squares) and
averaged (blue squares) column density of O{\sc vi} measurements
along the 91 LOS outlined in Fig. 1 (top).
It can be clearly seen in Fig.~\ref{FuseOegerle} that for $t=14.5$
Myr the calculated N(O{\sc vi}) distribution in the LB is similar to
that observed with FUSE. These results allow us to constrain the age
of the LB to be $14.5\pm^{0.7}_{0.4}$ Myr, and capture at the same
time all the relevant LB properties, such as size, age, dynamical
instabilities in the interaction shell, amongst others.

Successive blast waves advance very fast through the low-density
cavity, but slow down considerably later, as they run into the dense
shell. As a consequence an asymmetric reverse shock leads to shear
flow and turbulence inside the cavity, with a largest eddy size of a
fraction of the bubble diameter, typically $l
\leq 75$ pc (Avillez \& Breitschwerdt 2006). Thus the turnover time
scale is about $\tau_m \sim l/c_s \leq 3.7 \times 10^5$ yr, for an
averaged sound speed of $c_s \approx 200$ km/s after a SN explosion.
As the last SN occurred about $0.5$ Myr ago, we expect that SN
ejected oxygen inside the LB at present time should have a rather
uniform O/H distribution, a result that is also confirmed by FUSE
observations of the O/H ratio (Moos et al. 2002).

\section{Discussion and Conclusions}
\label{disc}
According to Cox (2004) a crucial test for any viable LB model is the
reproduction of the measured amount of O{\sc vi}. By embedding the LB
into a typical inhomogeneous ISM, we obtain bubbles, which have more
internal structure, exhibit on average lower pressure, consistent with
that observed in the local clouds, show substantial mass loading
(cf. Dyson \& Hartquist 1987), and are more susceptible to break-up of
the surrounding shell (both in the disk and perpendicular to it), than
was reported by previous authors. In fact there is some evidence from
stellar absorption line studies towards high latitude stars, that the
LB might be a Local Chimney (Welsh et al. 1999). Break-out of gas
flows from superbubbles might also be a clue to the solution of the
``energy problem'' observed in LMC bubbles (Oey \& Garc\'ia-Segura,
2004), which seem to be systematically too small compared to
similarity solutions.

Finally, we would like to stress that our LB evolution model has the
highest resolution so far obtained, showing structures down to 1.25
pc. We have repeated our runs with 0.625 pc finest resolution to
confirm that our results are \emph{not resolution dependent}. This
will be a prerequisite to model in a next step also the EUV and soft
X-ray emission, including the full non-equilibrium ionisation
structure. These calculations are underway.

\begin{acknowledgements}
  This work has been partially funded by FCT under
  PESO/P/PRO/40149/2000 to MAdeA and DB. We thank the two referees,
  Don Cox and an anonymous one, for their constructive criticism
  and helpful comments.

\end{acknowledgements}

\end{document}